\documentclass[prl,nofootinbib,superscriptaddress,twocolumn]{revtex4}
\usepackage[a4paper,left=1.5cm,right=1.5cm,top=3cm,bottom=3cm]{geometry}

\usepackage{amssymb,amsmath,amsfonts}
\usepackage[dvipsnames]{xcolor}
\usepackage{graphicx}
\usepackage{longtable}
\usepackage{verbatim}
\usepackage{color}

\usepackage{color}
\usepackage{hyperref}
\hypersetup{colorlinks, citecolor=bluscuro, linkcolor=black, urlcolor=bluscuro}
\definecolor{rossos}{cmyk}{0,1,1,0.55}
\definecolor{bluscuro}{rgb}{0.15, 0.2, .85}
\definecolor{bluchiaro}{cmyk}{1,.3,0.,0.1}

\graphicspath{{./Figures/}}
\def\bma#1{\mbox{\boldmath{$#1$}}}

\newcommand{\be}{\begin{equation}}
\newcommand{\ee}{\end{equation}}
\newcommand{\bea}{\begin{eqnarray}}
\newcommand{\eea}{\end{eqnarray}}
\newcommand{\beq}{\begin{equation}}
\newcommand{\eeq}{\end{equation}}
\def\beqa{\begin{eqnarray}}

\def\eeqa{\end{eqnarray}}

\def\lsim{\mathrel{\rlap{\lower4pt\hbox{\hskip0.5pt$\sim$}}
    \raise1pt\hbox{$<$}}}         
\def\gsim{\mathrel{\rlap{\lower4pt\hbox{\hskip0.5pt$\sim$}}
    \raise1pt\hbox{$>$}}}         

\usepackage[normalem]{ulem}

\newcommand{\arXiv}[2]{\href{http://arxiv.org/pdf/#1}{{\tt [#2/#1]}}}
\newcommand{\arXivold}[1]{\href{http://arxiv.org/pdf/#1}{{\tt [#1]}}}

\begin{document}

\title{A Fresh Look at the Calculation 
of Tunneling Actions}
\author{J.~R.~Espinosa }
\address{Institut de F\'{\i}sica d'Altes Energies (IFAE), The Barcelona Institute of Science and Technology (BIST),
Campus UAB, 08193 Bellaterra, Barcelona, Spain}
\address{ICREA, Instituci\'o Catalana de Recerca i Estudis Avan\c{c}ats, 
08010 Barcelona, Spain}
\address{Instituto de F\'{\i}sica Te\'orica UAM/CSIC,
Universidad Aut\'onoma de Madrid, 28049, Madrid, Spain}

\date{\today}

\begin{abstract}
\noindent
An alternative approach to the calculation of tunneling actions,  that control the exponential suppression of the decay of metastable phases, is presented. The new method circumvents the use of bounces in Euclidean space by introducing an auxiliary function, a tunneling potential $V_t$ that connects smoothly the metastable and stable phases of the field potential $V$. The tunneling action is obtained as the integral in field space of an action density that is a simple function of $V_t$ and $V$. This compact expression can be considered as a generalization of the thin-wall action to arbitrary potentials and allows a fast numerical evaluation with a precision below the percent level for typical potentials. The method can also be used to generate potentials with analytic tunneling solutions.
\end{abstract}

\maketitle


\paragraph{\bf $\bma{ \S\, 1}$ Introduction \label{sec:intro}} 

The decay of metastable phases via quantum or thermal fluctuations is a topic that appears often in many areas of physics, from condensed matter to cosmology. Such decays proceed by fluctuations that nucleate bubbles of the energetically preferred phase which are large enough to grow and eat out the metastable phase. For long-lived states, the decay rate (per unit volume) $\Gamma/V$ is exponentially suppressed. For the decay by quantum fluctuations $\Gamma/V=A\ e^{-S_{E}/\hbar}$, where $S_{E}$ is a 4-dimensional Euclidean tunneling action.

The calculation of these tunneling actions, crucial to estimate the lifetime of  metastable states, is a well understood problem (at least at the semi-classical level) that is quickly reviewed in $\S 2$. 
For a single scalar field (that corresponds to the order parameter
of the system), $S_E$ is the Euclidean action for a bubble configuration that connects the two phases. This instanton or bounce configuration is obtained by solving a nonlinear differential equation, the Euler-Lagrange Euclidean equation of motion with appropriate boundary conditions, see $\S 2$.

This Letter presents (in $\S 3$) an alternative route for computing tunneling actions that circumvents the use of a bounce solving instead for a tunneling potential, $V_t(\phi)$, that connects the metastable and stable phases. In this new language the tunneling action is given by a simple integral in field space [Eq.~(\ref{newSE})] of an action density that depends on the potential $V(\phi)$ and on $V_t(\phi)$.  This expression amounts to a generalization of the usual thin-wall approximation, but is valid for arbitrary potentials. 

One useful application of this novel approach is the estimate of the tunneling action by making simple approximations for the tunneling potential $V_t(\phi)$. As is demonstrated in $\S 4$ for two families of representative potentials, this estimate can be better than 1\%. The procedure is flexible
[depending on how sophisticated one is in approximating $V_t(\phi)$]
and very fast (relying on a simple integral and its minimization). 
A second application of the approach, detailed in $\S 5$, is to generate in a straightforward manner potentials $V(\phi)$ that allow for analytic solutions of the tunneling problem. 
The method can be extended to tunneling by thermal fluctuations, as shown in $\S 6$. In $\S 7$ we conclude with some comments on further possible  extensions and applications of this approach.

\begin{figure}[t!]
\includegraphics[width=0.9\columnwidth,height=0.6\columnwidth]{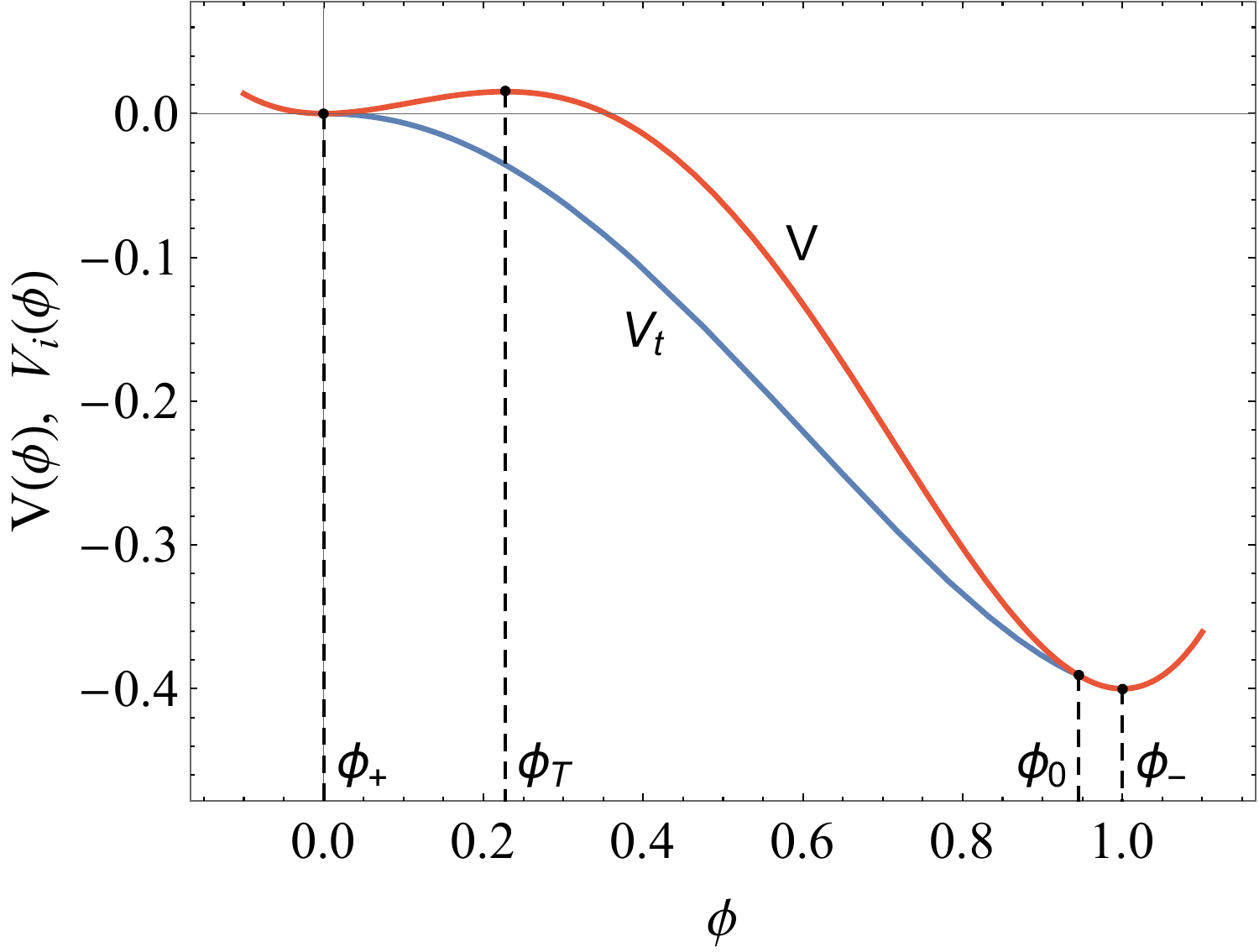}
\caption{\em Tunneling potential $V_t(\phi)$ (blue) for the
potential $V(\phi)$ (red) of Eq.~(\ref{V234}), with $\omega=0.4$.} 
\label{fig:VVt}
\end{figure}

\paragraph{\bf $\bma{\S\, 2}$ Euclidean Action via the Tunneling Bounce \label{sec:bounce}} 

Consider a scalar field $\phi$ in 4 dimensions, with a potential (or free-energy) $V(\phi)$ that has a metastable local minimum at $\phi_+$ and a deeper minimum at $\phi_-$, see Fig.~\ref{fig:VVt}. We set $\phi_+=0$ and $V(\phi_+)=0$ without loss of generality. 

The tunneling action for the decay of the metastable $\phi_+$ (false) vacuum is obtained \cite{Coleman} by finding an $O(4)$- symmetric bounce $\phi_b(r)$ (or Euclidean bubble) that interpolates between the false vacuum and (the basin of) the true vacuum at $\phi_-$. This bounce solves the  Euclidean equation of motion 
\be
\ddot{\phi} +\frac{3}{r}\dot{\phi} = V'\ ,
\label{EoM4}
\ee
where a dot (prime) represents a derivative with respect to $r$ ($\phi$), 
with the boundary conditions
\be
\dot\phi_b(0)=0\ ,\quad \phi_b(\infty)=\phi_+\ .
\ee
Identifying $r$ with time, Eq.~(\ref{EoM4}) corresponds to the classical motion of a particle in the inverted potential $-V(\phi)$ with a velocity and time dependent friction force. The solution can be found by undershooting and overshooting, changing the value of the field at the center of the Euclidean bubble, $\phi_b(r=0)\equiv\phi_0$, till the boundary condition at $r\rightarrow \infty$ is satisfied. The bounce $\phi_b$
extremizes \cite{CGM} the Euclidean action 
\bea
S_E[\phi_b] &=& 2\pi^2\int_0^\infty  \left[\frac12 \dot{\phi}_b^2 + V(\phi_b)-V(\phi_+)\right]r^3dr\nonumber\\
& = & S_{E,K}+S_{E,V}\ ,
\label{SE}
\eea
that gives the exponential suppression for the $\phi_+$ decay rate.
In the second line we have split the action in a gradient piece (from $\dot\phi_b^2/2$) plus a potential piece. Simple scaling arguments give \cite{Derrick}
\be
 S_{E,K}=2S_E\ ,\quad S_{E,V}=-S_E\ .
\label{Derrick}
\ee
When the potential difference between the vacua is very small, the bounce has a sharp transition between $\phi_0\simeq \phi_-$ for
$r<R_c$ and $\phi_+$ for $r>R_c$, where the bubble radius $R_c$
can be calculated analytically in terms of the surface tension 
\be
\sigma\simeq \int_{\phi_+}^{\phi_-}\sqrt{2[V(\phi)-V(\phi_-)]}d\phi\ ,
\ee 
as $R_c=3\sigma/\Delta V$, where $\Delta V\equiv V(\phi_+)-V(\phi_-)$. In this so-called ``thin-wall'' case the tunneling action can be obtained analytically \cite{Coleman} as
\be
S_{E,tw}=\frac{27\pi^2}{2}\frac{\sigma^4}{\Delta V^3}\ .
\label{SEtw}
\ee

The numerical calculation of $S_E$ can be cumbersome at times, especially for nearly-degenerate vacua, when finding $\phi_0$ might require high precision or the use of double shooting techniques \cite{Double}. Although the thin-wall result is applicable precisely in the near-degenerate case $\Delta V\rightarrow 0$, the approximation degrades quickly \cite{SH} with increasing $\Delta V$. At other times, finding the dependence of $S_E$ on model parameters would be useful. For a combination of these reasons, there have been many approaches in the literature to simplify the calculation of $S_E$. These include setting
bounds on $S_E$ \cite{Bounds};  improving the thin-wall approximation \cite{SH}; finding potentials that allow analytical solutions for the bounce and using piece-wise approximations to $V(\phi)$ based on such solutions \cite{exact}; perturbative expansions in $\Delta V$ \cite{Munster} or iterative methods \cite{Buniy}; numerical fits \cite{Adams}; etc.

\paragraph{\bf $\bma{\S\, 3}$ Euclidean Action via a Tunneling Potential \label{sec:Vint}} 

The new approach presented in this Letter has as central quantity an auxiliary function, $V_t(\phi)$, the ``tunneling potential''. In the language of the bounce discussed in the previous section, it is defined as
\be
V_t(\phi)\equiv V(\phi) -\frac12 \dot\phi_b^2\ ,
\label{Vt}
\ee
where it is understood that $\dot\phi_b$ above should be expressed in terms of the field $\phi$.
 
Some properties of $V_t(\phi)$ are the following:
1) $V_t(\phi)$ is a monotonic function, with $V_t(\phi)\leq V(\phi)$. Notice 
that $V_t(\phi)$ is just minus the Euclidean energy. As the Euclidean energy is dissipated by the friction term in (\ref{EoM4}):
\be
\frac{d}{dr}\left[\frac12 \dot\phi_b^2-V(\phi_b)\right]=-\frac{3}{r}\dot\phi_b^2\leq 0\ ,
\ee
it decreases monotonically with $r$. Noting that the bounce is also a monotonic function of $r$ \cite{CGM} (intuitively clear from the ``motion in an inverted potential'' picture), monotonicity of $V_t(\phi)$ follows.  
2) $V_t(\phi)$ is defined in the interval between $\phi_+$ and $\phi_0$ only [the mapping in $\phi$ of the infinite interval $r\in (0,\infty)$]. At these end points, $V_t=V$, as $\dot\phi_b(0,\infty)=0$. 
An example of $V_t(\phi)$ is shown in Fig.~\ref{fig:VVt}.

We can remove altogether the reference to the bounce (and the 
4-dimensional Euclidean space in which it lives) in favor of $V_t(\phi)$ as follows. From (\ref{Vt})
\be
\dot\phi_b = - \sqrt{2[V(\phi)-V_t(\phi)]}\ ,
\label{dphi}
\ee
where the minus sign, chosen due to $\phi_+<\phi_-$, should be a plus when $\phi_+>\phi_-$. Eq.~(\ref{dphi}) allows to remove any derivative of $\phi_b$ in terms of $V_t$ (and $V$). The Euclidean
radial coordinate $r$ can be extracted from (\ref{EoM4})
as
\be
r=3\sqrt{2(V-V_t)/(V_t')^2}\ .
\label{r}
\ee
Taking a derivative of the above with respect to $r$ we arrive at a differential equation for $V_t$:
\be
\boxed{\left(4V_t'-3 V' \right)V_t' = 6(V_t-V)V_t''}\ ,
\label{VtEoM}
\ee
which takes the place of (\ref{EoM4}) in the new formulation of the tunneling problem: find a $\phi_0$ and a $V_t(\phi)$ that solve (\ref{VtEoM}) with the boundary conditions:
\be
V_t(\phi_+)=V(\phi_+)\ ,\quad V_t(\phi_0)=V(\phi_0)\ .
\label{BCV}
\ee
Eq.~(\ref{VtEoM}) also leads [assuming $V'(\phi_+)=0$] to
\be
V'_t(\phi_+)=0\ ,\quad V'_t(\phi_0)=3 V'(\phi_0)/4\ .
\label{BCVp}
\ee
The tunneling action (\ref{SE}) can also be rewritten in terms of $V_t(\phi)$. It is convenient, for reasons to be discussed below, to use $S_E=S_{E,K}/2$ [see (\ref{Derrick})] to get
\be
\boxed{S_E[V_t]=54 \pi^2\int_{\phi_0}^{\phi_+}\frac{(V-V_t)^2}{\left(V_t'\right)^3}d\phi}
\label{newSE}
\ee
The new differential Eq.~(\ref{VtEoM}) seems hardly an improvement over the original formulation of Eq.~(\ref{EoM4}) and, in particular, it cannot be linearized, see {\it e.g.} \cite{Ibragimov}. However, it turns out to be quite simple to estimate $V_t(\phi)$ for a given potential to get an accurate approximation to the tunneling action using (\ref{newSE}), as shown in $\S 4$.

Equation (\ref{newSE}) can be regarded as a generalization of the thin-wall approximation that is valid for generic potentials. In the limit of quasi-degenerate vacua we recover the
thin-wall result as follows. Given the properties of $V_t$, for near degenerate vacua one has $V_t'\ll (V-V_t)'$, and (\ref{VtEoM})
can be reduced to the form $(V_t-V)'V_t'\simeq 2(V_t-V)V_t''$.
This can be readily integrated to get 
\be
\sqrt{V-V_t}\simeq \kappa V_t'\ ,
\label{tw}
\ee 
with $\kappa$ an integration constant. Integrating (\ref{tw}) gives
\be
\int_{\phi_+}^{\phi_-}\sqrt{V(\phi)-V_t(\phi)}d\phi \equiv \sigma/\sqrt{2}=-\kappa \Delta V\ .
\label{sigmatw}
\ee
Plugging (\ref{tw}) and (\ref{sigmatw}) in the action (\ref{newSE})
reproduces the thin-wall result (\ref{SEtw}).

To sum up, $V_t(\phi)$ captures all the relevant information of the tunneling bounce and has the appealing property of being quite featureless [{\it e.g.} a potential with a bumpy barrier gives a $\dot\phi^2/2$ that also has bumps. Both sets of bumps cancel out in $V_t(\phi)$]. This property can be exploited to estimate the tunneling action as is shown next.

\paragraph{\bf $\bma{\S\, 4}$ Estimates of the Euclidean Action\label{sec:SEestimate}} 
Let us give approximations for $V_t$ of increasing level of sophistication. The simplest is the linear one:
\be
V_{t1}(\phi)=V_0\frac{\phi}{\phi_0}\ ,
\label{Vi1}
\ee
where $V_0=V(\phi_0)$, and it satisfies the boundary conditions (\ref{BCV}). To also satisfy the boundary condition (\ref{BCVp}) on $V_t'(\phi_0)$, one can use the quadratic approximation
\be
V_{t2}(\phi)=V_{t1}(\phi)+\frac{\phi}{4\phi_0^2}\left(3\phi_0V'_0-4V_0\right)(\phi-\phi_0)\ ,
\label{Vi2}
\ee
where $V'_0=V'(\phi_0)$. One can go further, matching also $V_t'(\phi_+)$ using the cubic approximation
\be
V_{t3}(\phi)=V_{t2}(\phi)+\frac{\phi}{4\phi_0^3}\left(3\phi_0V'_0-8V_0\right)(\phi-\phi_0)^2\ .
\label{Vi3}
\ee
So far, the previous approximations focus only on the boundary conditions at $\phi_+$ and $\phi_0$. We can add information on 
the top of the barrier separating the vacua, forcing the approximation to satisfy Eq.~(\ref{VtEoM}) at $\phi_T$, the value at which $V(\phi)$ is maximal. This can be achieved with the quartic approximation 
\be
V_{t4}(\phi)=V_{t3}(\phi)+a_4 \phi^2(\phi-\phi_0)^2\ ,\label{Vi4}
\ee
where
\be
a_4=\frac{1}{c}\left(
a_{0T}-\sqrt{a_{0T}^2-c\, U_{t3T}}\right)\ ,
\ee
with 
$
c\equiv 4\phi_T^2\phi_{0T}^2(\phi_0^2+2\phi_{0T}\phi_T)
$, 
$\phi_{0T}\equiv \phi_0-\phi_T$,
$U_{t3T}\equiv 4(V'_{t3T})^2+6(V_T-V_{t3T})V''_{t3T}$,
 $V_{t3T}\equiv V_{t3}(\phi_T)$ and
\bea
a_{0T}&=&
-6(V_T-V_{t3T})(\phi_0^2-6\phi_{0T}\phi_T)\\
&&-8\phi_T(\phi_{0T}-\phi_T)\phi_{0T}V'_{t3T}
+3\phi_T^2\phi_{0T}^2V''_{t3T} .\nonumber
\eea
This approximation also satisfies the boundary conditions in Eqs. (\ref{BCV}) and (\ref{BCVp}). 

 One virtue of the expression we have chosen for $S_E$
is that it is stationary with respect to variations of $V_t$. That is,
$\delta S_E/\delta V_t=0$ gives back (\ref{VtEoM}). Moreover, as proven in the Appendix, the true $V_t$ minimizes the action. Therefore,
for an approximate ansatz $V_{ta}$,  (\ref{newSE}) gives an approximate action $S_{E,a}$ bigger than the true $S_E$. $S_{E,a}$ is a function of the end point $\phi_0$ and its minimum gives the best estimate for $S_E$. To illustrate this and check the efficiency of the approximations above let us 
consider two representative potentials, already used in \cite{SH}
to improve the thin-wall approximation.

Both potentials are rescaled so that $\phi$ is dimensionless and  have the false vacuum at $\phi_+=0$, $V(\phi_+)=0$ and the true vacuum at $\phi_-=1$, $V(\phi_-)=-\omega$. The parameter $\omega$ controls the energy difference between both vacua. The first potential is a quartic potential
\be
V(\phi)= \phi^2-2(1+2\omega)\phi^3+(1+3\omega)\phi^4\ .
\label{V234}
\ee
One example, for $\omega=0.4$, is shown in Fig.~\ref{fig:VVt}.
The second potential, constructed to allow for higher barriers
when $\omega$ is large, reads, with $a=\log(1+\omega)/(2\pi)$,
\be
V(\phi) = 1-\exp(2\pi a \phi)\left[\cos(2\pi\phi)-a\sin(2\pi\phi)\right]\ .\label{Vexp}
\ee
\begin{figure}
\includegraphics[width=0.9\columnwidth,height=0.6\columnwidth]{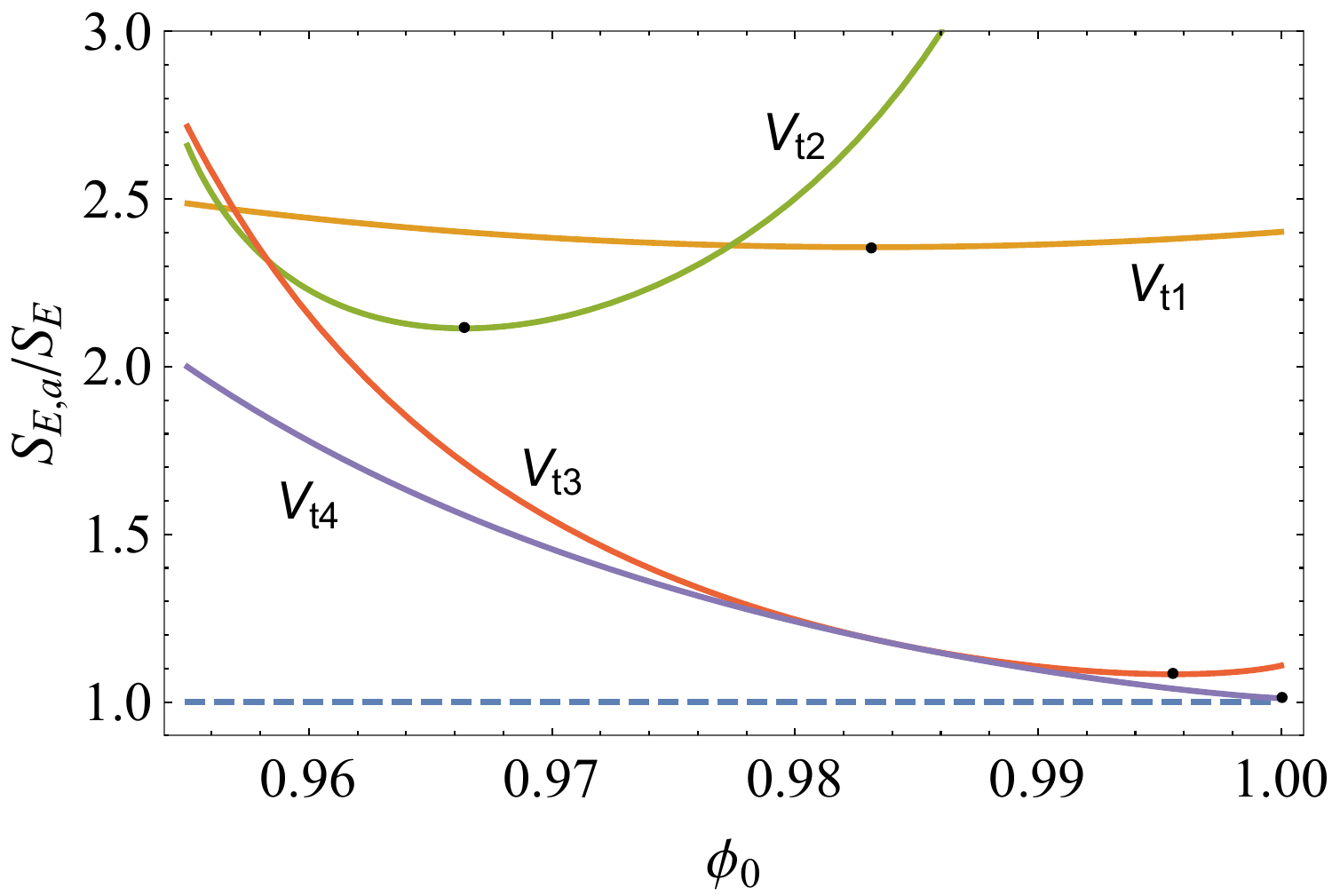}
\caption{\em Estimates $S_{E,a}$ of the tunneling action (normalized to the exact $S_E$) 
 for the potential
  (\ref{Vexp})  as a function of  $\phi_0$,  for $\omega=2$, corresponding to the approximations  (\ref{Vi1})-(\ref{Vi4}) to the tunneling potential. Dots mark the  minima as  the best estimate for $S_{E}$ in each case.}
\label{fig:SEa}
\end{figure}
The approximate actions $S_{E,a}(\phi_0)$ for the four different $V_{ta}$ approximations introduced above are shown (normalized to the exact action $S_E$) in Fig.~\ref{fig:SEa}, for the potential  (\ref{Vexp}) taking $\omega=2$ [the result is similar for potential (\ref{V234})]. The best estimate in each case corresponds to the marked minimum. While $V_{t1}$ gives an order of magnitude accuracy,
$V_{t4}$ is accurate below the $\%$ level. Convergence of successive approximations can also be used to gauge how close one is to the true result.

\begin{figure}
\includegraphics[width=0.9\columnwidth,height=0.6\columnwidth]{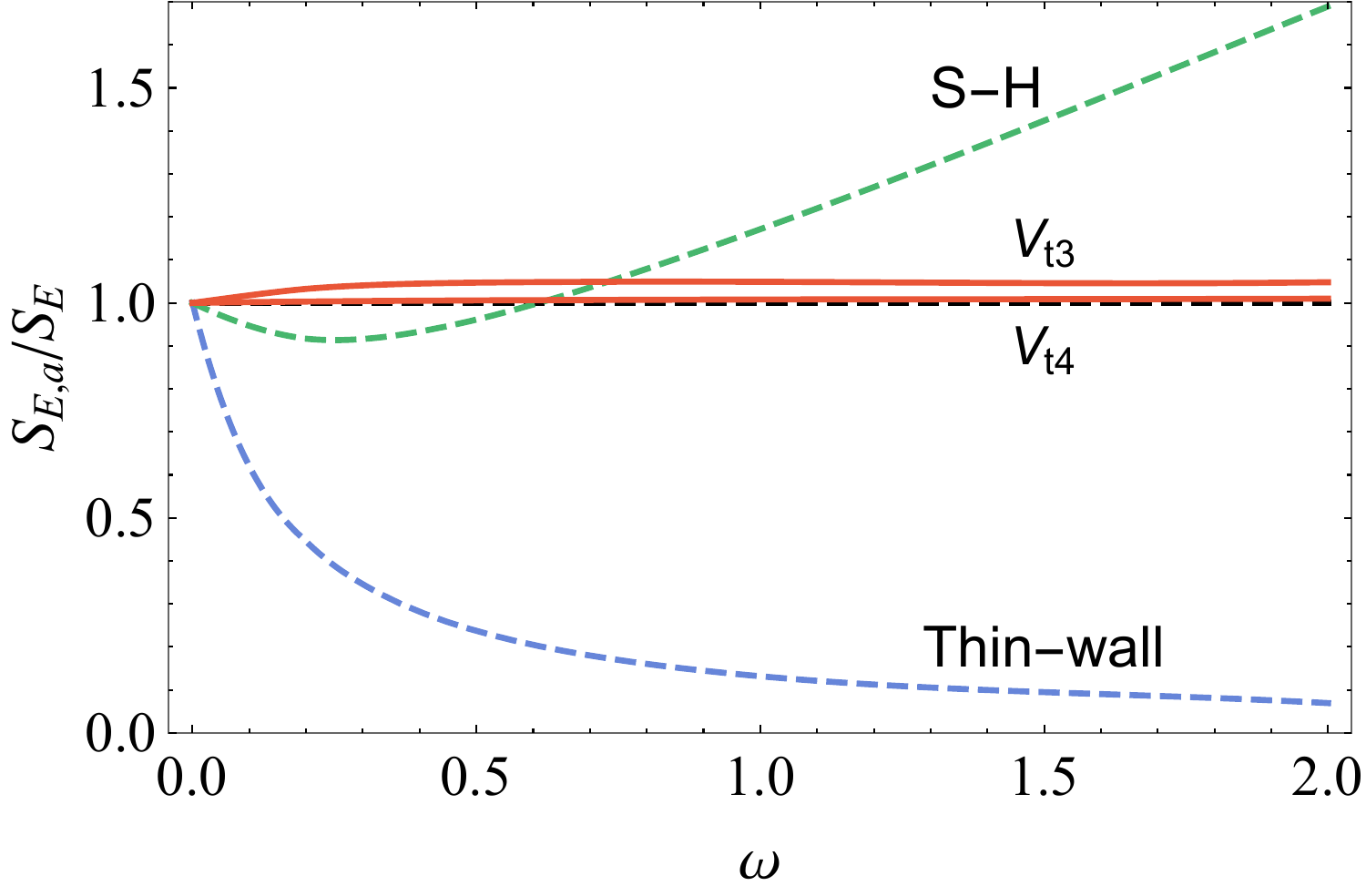}
\includegraphics[width=0.9\columnwidth,height=0.6\columnwidth]{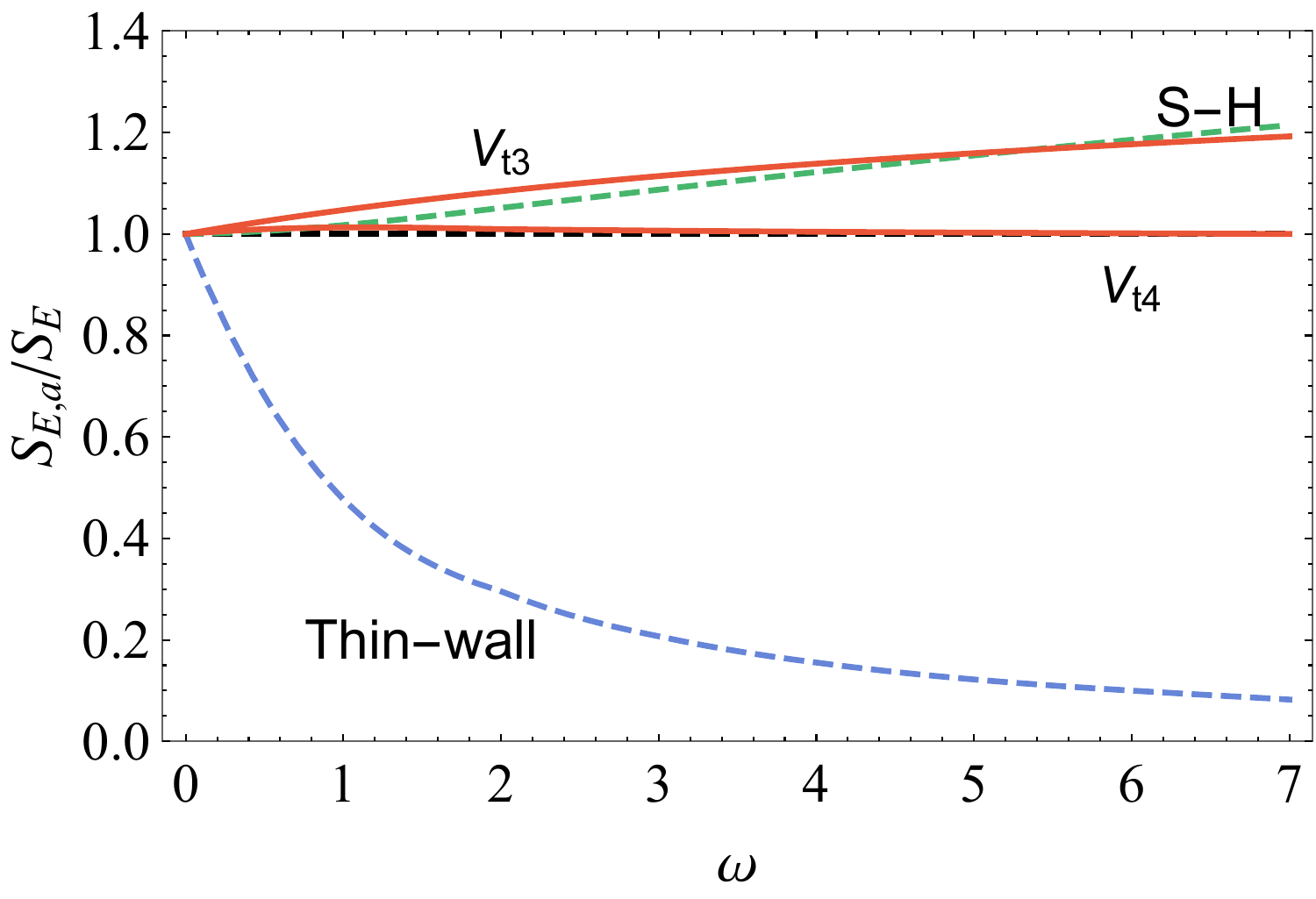}
\caption{\em Estimates $S_{E,a}$ of the tunneling action (normalized to the exact action $S_E$)  for the potentials (\ref{V234}) (upper plot) and (\ref{Vexp}) (lower plot)   as a function of $\omega$, that controls the energy difference between the vacua: the thin-wall estimates (\ref{SEtw}); the estimate in \cite{SH} (label ``S-H''); cubic  (\ref{Vi3}) and quartic (\ref{Vi4})
 approximations to the tunneling potential.}
\label{fig:comp}
\end{figure}

The ratio $S_{E,a}/S_E$ as a function of $\omega$ for the most accurate estimates, from $V_{t3}$ and $V_{t4}$, is shown,
for both potentials (\ref{V234}) and (\ref{Vexp}), in Fig.~\ref{fig:comp}.  For comparison, the figure also shows the thin-wall approximation (accurate only very close to $\omega=0$) and the improved approximation (labelled ``S-H''), both as calculated in \cite{SH}. Once again we see that $V_{t4}$ leads to an excellent approximation of the tunneling action, at the level of $1\%$
or below. 

\paragraph{\bf $\bma{\S\, 5}$ Potentials with Exact Tunneling Solutions\label{sec:Exact}} 
Another application of the new approach to tunneling action calculations based on (\ref{VtEoM}) is that it allows to generate
in a straightforward manner potentials (or pieces of them) that 
lead to analytic solutions of the tunneling problem. Instead of
starting from $V$ and solving for $V_t$, one postulates a given 
$V_t$ and  integrates (\ref{VtEoM}) to obtain the corresponding $V$ as
\be
\frac{V(\phi)}{\left[V'_t(\phi)\right]^2}=
\frac{V_{t,0}}{\left(V'_{t,0}\right)^2}+
\int_{\phi_0}^\phi\frac{
4\left[V'_t(\bar\phi)\right]^2-6V_t(\bar\phi)V''_t(\bar\phi)}{3\left[V'_t(\bar\phi)\right]^3}\,d\bar\phi\ ,
\label{VfromVt}
\ee
where $V_{t,0}=V_t(\phi_0)$, etc. As a trivial example, $V_t(\phi)=a\phi^3$, leads to $V=a\phi^4/\phi_0$, a well known case of potential that allows for an analytic solution. Setting $a=-\lambda\phi_0/4$, and plugging $V$ and $V_t$ in the action integral (\ref{newSE}), we recover the known action $S_E=8\pi^2/(3\lambda)$ for $V=-\lambda\phi^4/4$, a result that is relevant for the decay of the electroweak vacuum in the Standard Model \cite{STAB}. 

Among many other possible examples, consider
\be
V_t(\phi) = -A \tanh(B \phi^2)\ ,
\ee
where $A$ and $B$ are constants that control the depth and slope of $V_t(\phi)$. Using (\ref{VfromVt}) the corresponding potential is
\bea
V(\phi) &=& -A\left\{\frac{B\phi^2}{6\ {\cosh^4(B\phi^2)}}\right[{\rm Chi}(2B\phi^2)-{\rm Chi}(2B\phi_0^2)
\nonumber\\
&&+\left.\left.\log\frac{\phi^2}{\phi_0^2}\right]
+\tanh(B\phi^2)\right\}\ ,
\eea
where ${\rm Chi}(x)$ is the hyperbolic cosine integral function.

Although in the previous example the tunneling action cannot be obtained in an analytic closed form, one can easily find examples in which that is possible. For instance, from
\be
V_t(\phi)=A\,\phi^2(2\phi-3)\ ,
\ee
one gets
\be
V(\phi)=A\,\phi^2\left[2\phi-3+2(1-\phi)^2\log
\frac{(1-\phi)\phi_0}{(1-\phi_0)\phi}
\right]\ .
\ee
The tunneling action is 
\bea
S_E&=&-\frac{\pi^2}{3A}\left[\phi_0+{\rm Li}_2\left(\frac{\phi_0}{\phi_0-1}\right)\right]\ .
\eea
The profile of the tunneling bubble can also be obtained as the simple function
\be
\phi_b(r)=\frac12\left[1-\tanh(A\, r^2/2-B) \right]\ ,
\ee
with $B=-\tanh^{-1}(1-2\phi_0)$, so that $\phi_b(0)=\phi_0$. 
This case provides a complete analytical solution with a fully differentiable potential that admits also a thin-wall limit ($B\gg 1$). 
In both examples, $\phi$ is dimensionless and normalized so that the minimum is at $\phi=1$.

When an approximate ansatz for $V_t$ is used to approximate  the tunneling rate for a given potential $V(\phi)$, formula (\ref{VfromVt}) can also be used to check how close is the potential derived from the ansatz  to the original potential.


\paragraph{\bf $\bma{\S\, 6}$ Thermal Tunneling\label{sec:thermal}} 
For the decay of a metastable phase by thermal fluctuations \cite{Langer} the decay rate per unit volume is $\Gamma/V=A_T\ e^{-S_{E3}/T}$ where $S_{E3}$ is a 3-dimensional Euclidean tunneling action and $T$ is the temperature of the system. The previous analysis can be carried out in a similar manner simply changing the dimension of the Euclidean space to $d=3$ and taking into account that the potential $V(\phi)$ receives thermal corrections. Keeping the Euclidean dimension $d$ unspecified, the $O(d)$-invariant bounce is the solution to the differential equation
\be
\ddot\phi +\frac{d-1}{r}\dot\phi = V'\ ,
\label{EoMd}
\ee  
with the same boundary conditions (\ref{BCV}). The tunneling potential $V_t$ is defined as in (\ref{Vt}) and the bounce can be removed as before. From (\ref{EoMd}) one gets the relation 
\be
r=(d-1)\sqrt{2(V-V_t)/(V_t')^2}\ ,
\ee
and its derivative gives the differential equation for $V_t$:
\be
(V_t')^2=\frac{(d-1)}{d}\left[V' V_t' - 2(V_t-V)V_t''\right]\ .
\ee
In a similar manner the tunneling action is
\bea
S_{E,d}&=&\frac{2\pi^{d/2}}{\Gamma(d/2)}\int_0^{\infty}\left[\frac12\dot\phi^2+V(\phi)-V(\phi_+)\right]r^{d-1}dr\nonumber\\
&=&S_{E,d,K}+S_{E,d,V}\ ,
\eea
with the relations
\be
S_{E,d,K}=\frac{d}{2}S_{E,d}\ ,\quad
S_{E,d,V}=\frac{2-d}{2}S_{E,d}\ .
\ee
In terms of $V$ and $V_t$ this action can be written as
\be
S_{E,d}=\frac{(d-1)^{(d-1)}(2\pi)^{d/2}}{\Gamma(1+d/2)}\int_{\phi_+}^{\phi_0}\frac{(V-V_t)^{d/2}}{\left|V_t'\right|^{d-1}}d\phi\ .
\ee
The case of thermal decay simply corresponds to $d=3$.

\paragraph{\bf $\bma{\S\, 7}$ Outlook\label{sec:out}} 
The new approach to the calculation of tunneling actions presented in this Letter can be generalized or extended in several ways:

First, the method is applicable to tunneling in multi-field potentials and it offers significant advantages in estimating the tunneling action \cite{multi} as one is now searching for a minimum rather than a saddle point.

Second, in applications to cosmology, gravitational effects play a fundamental role \cite{CdL} and should be included. It is straightforward to generalize the differential equation (\ref{VtEoM}) for the tunneling potential $V_t$ to include such gravitational corrections, obtaining
\be
\left(4V_t'-3 V' \right)V_t' = 6(V_t-V)\left(V_t''+\frac{3V-2V_t}{m_P^2}\right)
\ ,
\label{VtEoMgrav}
\ee
where $m_P=2.435\times 10^{18}$ GeV is the reduced Planck mass. The equation above reduces to the case without gravity for
$m_P\rightarrow \infty$, as it should. The generalization of the tunneling action expressions in terms of $V_t$ in the gravitational case requires a dedicated study that goes beyond the scope of this Letter (see \cite{grav}).

Finally, similar approaches that get rid altogether of
an Euclidean space to find instanton solutions might also be applicable to study other semiclassical objects like sphalerons, etc.

These applications will be explored in future works.

\paragraph{\bf $\bma{\S\,}$ Appendix. Proof that $\bma{S_E[V_t]}$ is a minimum.\label{sec:out}} 

Let us write the action functional (\ref{newSE}) as
\be
S_E[V_t]=\int_{\phi_+}^{\phi_0}{\cal S}_E\,d\phi\ ,
\ee
with the (positive definite) action density
\be
{\cal S}_E={\cal S}_E(\phi,V_t,V_t')=
54 \pi^2\frac{(V-V_t)^2}{\left(-V_t'\right)^3}\ .
\ee
An elegant proof that $S_E[V_t]$
is a global minimum for the $V_t$ that solves the differential Eq.~(\ref{VtEoM}) with boundary conditions (\ref{BCV}) can be given via  the general method of `fields of extremals', see {\it e.g.} \cite{Russak}.

\begin{figure}
\includegraphics[width=0.9\columnwidth,height=0.6\columnwidth]{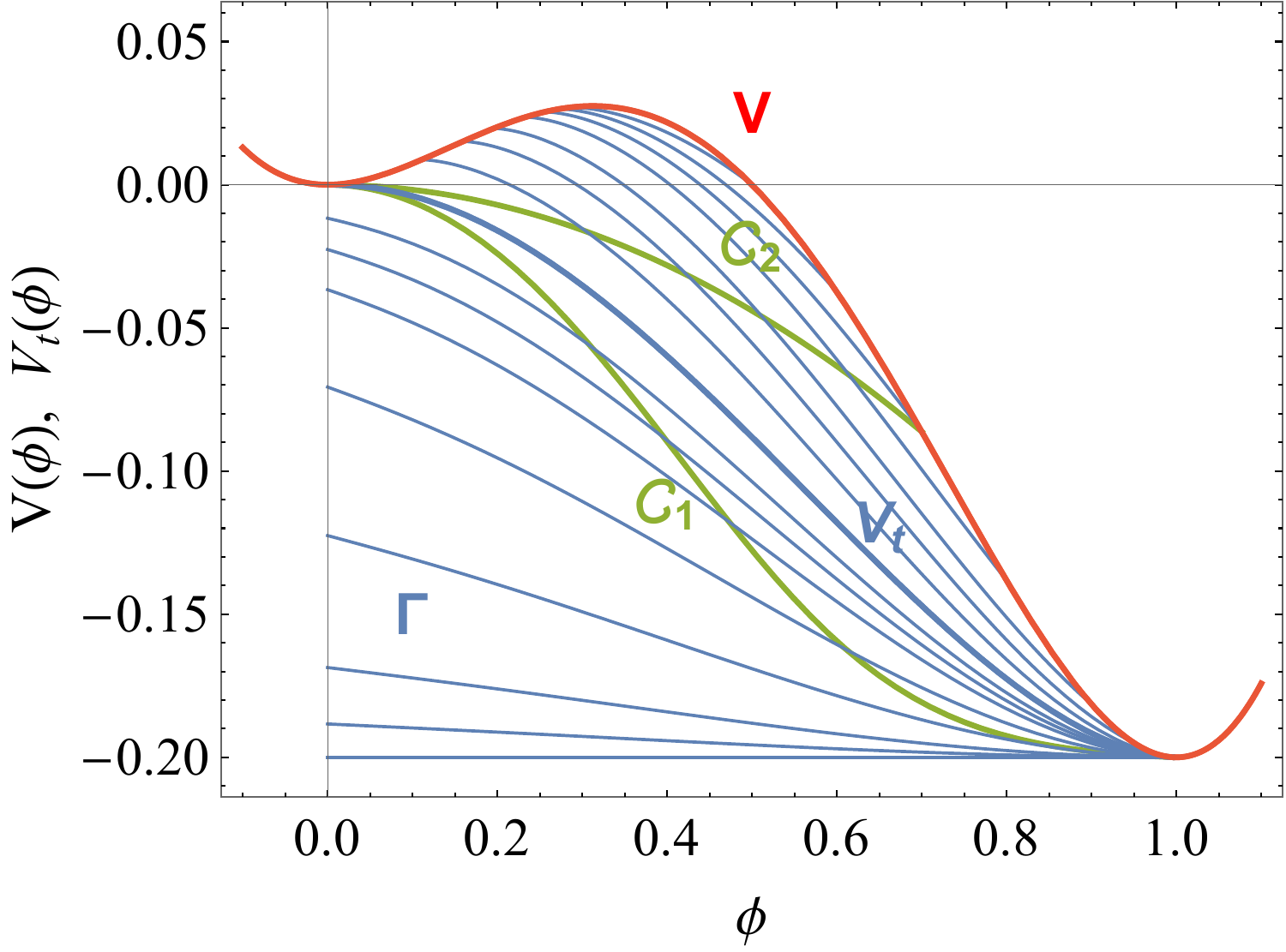}
\caption{\em In blue, different $V_t(\phi)$ functions corresponding to solutions of the Euler-Lagrange Eq.~(\ref{EoM4}) with different initial conditions $\phi(0)=\phi_i$ for the potential $V(\phi)$ in red. The thick blue line is the true tunneling solution for $\phi(0)=\phi_0$. Above (below) it there are undershooting (overshooting) solutions. The area covered, $\Gamma$, is a ``field of extremals''.  Two alternative paths, ${\cal C}_{1,2}$, out of the false minimum are shown in green.}
\label{fig:field}
\end{figure}

Consider the inverted potential approach to the tunneling problem.
Varying the initial value of the field $\phi(0)=\phi_i$ (with $\dot\phi(0)=0$) and solving the equation of motion (\ref{EoM4}) we find overshooting  and undershooting solutions (for $\phi_i>\phi_0$ and $\phi_i<\phi_0$ respectively). For each of these solutions we can define, via (\ref{Vt}), a function $V_t(\phi)$ (a generalization of the $V_t(\phi)$ for the true tunneling function, corresponding to $\phi=\phi_0$). We consider the evolution of the field only up to $\phi_+=0$ for overshoots and 
up to the point at which $\dot\phi=0$ for undershoots. In this way a one-parameter family of extremals [solutions to the Euler-Lagrange eq.~(\ref{EoM4})] is generated. This is shown in Fig.~\ref{fig:field}
for an example potential. The area of the $(\phi,v_t)$ plane covered by these solutions is called `field of extremals', labeled by $\Gamma$ in the plot. Through every point inside this field  $\Gamma$ a single extremal passes and defines a slope
\be
p(\phi,v_t)=V'_t(\phi)\ .
\ee
This slope function is used in the definition of Hilbert's invariant functional
\bea
&&\hspace*{-0.5cm}
\label{SEstar}
S_E^*[{\cal C}]\equiv\int_{\cal C}{\cal S}^*_E(\phi,v_t,p)\, d\phi\\
&&\equiv\int_{\cal C}\left[{\cal S}_E(\phi,v_t,p)\,d\phi
+(dv_t-p\,d\phi)\frac{\partial{\cal S}_E}{\partial v_t'}(\phi,v_t,p)\right] ,\nonumber
\eea
where ${\cal C}$ is any curve $v_t(\phi)$ inside $\Gamma$. We are interested in curves starting at 
the false minimum and ending on the right slope of $V$. The differentials $d\phi$ and $dv_t$ are understood to be taken along ${\cal C}$. Example curves are shown in Fig.~\ref{fig:field} in green.

Two interesting properties of $S_E^*$ are the following: 1)
along an extremal curve $v_t(\phi)$, $S_E^*=S_E$ as $p=dv_t/d\phi$ along such curves; 2) the value of $S_E^*$ depends only on
the location of the end points of the curve. This follows from the fact that the integrand of (\ref{SEstar}) is an exact differential.\footnote{For $V=-\lambda\phi^4/4$ the exact differential $dS_E^*$ can be obtained as $dS_E^*=-(8\pi^2/\lambda)\, d\left[a\left(a^2/3+a+1\right)\right]$
with $a=\lambda\phi^4/(4v_t)$.}  Indeed it is straightforward to show that 
\be
\frac{d}{d\phi} \frac{\partial{\cal S}_E}{\partial v_t'}=
\frac{d}{d v_t} \left[{\cal S}_E
-p\frac{\partial{\cal S}_E}{\partial v_t'}\right]\ ,
\ee
or, more explicitly,
\be
\frac{\partial^2{\cal S}_E}{\partial\phi\partial v_t'}+
\frac{\partial^2{\cal S}_E}{(\partial v_t')^2}\frac{\partial p}{\partial\phi}=
\frac{\partial{\cal S}_E}{\partial v_t}-p
\frac{\partial^2{\cal S}_E}{\partial v_t\partial v_t'}-p
\frac{\partial^2{\cal S}_E}{(\partial v_t')^2}
\frac{\partial p}{\partial v_t}\ ,
\ee
by
noting that the Euler-Lagrange equation of motion $d(\partial{\cal S}_E/\partial v_t')/d\phi=\partial{\cal S}_E/\partial v_t$ implies
\be
\frac{\partial {\cal S}_E}{\partial v_t} = \frac{\partial^2{\cal S}_E}{\partial\phi\partial v_t'}+p\,\frac{\partial^2{\cal S}_E}{\partial v_t\partial v_t'}+\frac{dp}{d\phi}\,\frac{\partial^2{\cal S}_E}{(\partial v_t')^2}\ ,
\ee
with $dp/d\phi=\partial p/\partial\phi+p\, \partial p/\partial v_t$.
As a corollary of these two properties above, the value of $S_E^*$ along any curve (inside the field of extremals) connecting the false minimum and the  point $(\phi_0, V(\phi_0))$, like ${\cal C}_1$ in Fig.~\ref{fig:field}, is independent of the shape of the curve and equals $S_E[V_t]$. Noting further that ${\cal S}^*_E=0$ on the right slope of $V$, we conclude that $S_E^*$ is the same for any curve (inside $\Gamma$) that joins the metastable minimum and the right slope of $V$. Referring to Fig.~\ref{fig:field},  $S^*_E[{\cal C}_2]=S_E[V_t]$.

To complete the proof we use the explicit form of ${\cal S}^*_E$ to show that it satisfies the inequality
\be
{\cal S}_E^*=54\pi^2\frac{(V-v_t)^2}{(-v_t')^3}\left[4-3\frac{v_t'}{p}\right]\left(\frac{v_t'}{p}\right)^3\leq {\cal S}_E\ ,
\label{SEineq}
\ee
as $(4-3x)x^3\leq 1$ for any $x$, with the equality holding for $x=1$ only. Therefore, for any curve ${\cal C}$ from the false minimum to the right slope of $V$ we find
\be
S_E[{\cal C}] \geq S_E^*[{\cal C}] = S_E[V_t]\ ,
\ee
which proves that $S_E[V_t]$ indeed minimizes the action. The equality sign requires $x=v'_t/p=1$ which corresponds to the
curve $V_t$, so no other curve shares this minimum value.
 
 This minimum property of the tunneling functional also holds for the thermal case. For generic $d$, the proof goes through as before with 
 \be
{\cal S}_E^*={\cal S}_E\left[d-(d-1)\frac{v_t'}{p}\right]\left(\frac{v_t'}{p}\right)^{d-1}\ .
\label{SEineqd}
\ee
For $d=3$ we get ${\cal S}_E^*\leq {\cal S}_E$ thanks to $(3-2x)x^2\leq 1$ for $x\geq 0$. In this case, therefore the minimum of $S_E[V_t]$ holds in the class of monotonically decreasing functions $v_t(\phi)$, for which $v_t'(\phi)/p\geq 0$.

\begin{acknowledgments}
\paragraph{\bf Acknowledgments.} I thank G. Ballesteros, J.L.F. Barb\'on, J. Elias-Mir\'o, A. Casas, T. Konstandin, J. Moreno, D. Racco and A. Riotto for useful comments
and reading the manuscript.
Work supported by the ERC
grant 669668 -- NEO-NAT -- ERC-AdG-2014, the Spanish MINECO grants  2016-78022-P and
FPA2014-55613-P, the Severo Ochoa excellence program of MINECO (grants SEV-2016-0588 and SEV-2016-0597) and by the Generalitat de Catalunya grant 2014-SGR-1450. 
\end{acknowledgments}



\begin{thebibliography}{99}



\bibitem{Coleman}
  S.R.~Coleman,
  Phys.\ Rev.\ D {\bf 15} (1977) 2929
   Erratum: [Phys.\ Rev.\ D {\bf 16} (1977) 1248].

\bibitem{CGM}
  S.R.~Coleman, V.~Glaser and A.~Martin,
  Commun.\ Math.\ Phys.\  {\bf 58} (1978) 211.
  
\bibitem{Derrick}
  G.H.~Derrick,
  J.\ Math.\ Phys.\  {\bf 5} (1964) 1252.
    
  
\bibitem{Double}
A. Bayliss, 
Mathematics of Computation 32 (1978) 61;
  G.V. Dunne and H. Min,
  Phys.\ Rev.\ D {\bf 72} (2005) 125004
  \arXivold{hep-th/0511156}.
  

\bibitem{SH}
  D.A.~Samuel and W.A.~Hiscock,
  Phys.\ Lett.\ B {\bf 261} (1991) 251.


\bibitem{Bounds}
  R.~Sato and M.~Takimoto,
  \arXiv{1707.01099}{hep-ph};
  A.R.~Brown,
  \arXiv{1711.07712}{hep-th}.
  
\bibitem{exact}
  A.~Ferraz de Camargo, R.C.~Shellard and G.C.~Marques,
  Phys.\ Rev.\ D {\bf 29} (1984) 1147;
  K.M.~Lee and E.J.~Weinberg,
  Nucl.\ Phys.\ B {\bf 267} (1986) 181.
  M.J.~Duncan and L.G.~Jensen,
  Phys.\ Lett.\ B {\bf 291} (1992) 109;
  T.~Hamazaki, M.~Sasaki, T.~Tanaka and K.~Yamamoto,
  Phys.\ Rev.\ D {\bf 53} (1996) 2045
  \arXivold{gr-qc/9507006};
  K.~Dutta, C.~Hector, P.M.~Vaudrevange and A.~Westphal,
  Phys.\ Lett.\ B {\bf 708} (2012) 309
  \arXiv{1110.2380}{hep-th};
  K.~Dutta, C.~Hector, T.~Konstandin, P.M.~Vaudrevange and A.~Westphal,
  Phys.\ Rev.\ D {\bf 86} (2012) 123517
  \arXiv{1202.2721}{hep-th};
  G.~Pastras,
  JHEP {\bf 1308} (2013) 075
  \arXiv{1102.4567}{hep-th};
  A.~Aravind, B.S.~DiNunno, D.~Lorshbough and S.~Paban,
  Phys.\ Rev.\ D {\bf 91} (2015) 2,  025026
  \arXiv{1412.3160}{hep-th};
  V.~Guada, A.~Maiezza and M.~Nemevšek,
  \arXiv{1803.02227}{hep-th}.
    
\bibitem{Munster}
  G.~Munster and S.~Rotsch,
  Eur.\ Phys.\ J.\ C {\bf 12} (2000) 161
  \arXivold{cond-mat/9908246}.
      
      
\bibitem{Buniy}
  R.V. Buniy,
  \arXiv{1610.00018}{hep-th}.
  
 
\bibitem{Adams}
  F.C.~Adams,
  Phys.\ Rev.\ D {\bf 48} (1993) 2800
  \arXivold{hep-ph/9302321}.
 
\bibitem{Ibragimov} 
N.H. Ibragimov, ``A practical course in differential equations and mathematical modelling'', ALGA Publications, Bleking Institute of technology Karlskrona, Sweden,
2006.
 

 
\bibitem{STAB}
  N.~Cabibbo, L.~Maiani, G.~Parisi and R.~Petronzio,
  Nucl.\ Phys.\ B {\bf 158} (1979) 295;
  N.V.~Krasnikov,
  Yad.\ Fiz.\  {\bf 28} (1978) 549;
  P.~Q.~Hung,
  Phys.\ Rev.\ Lett.\  {\bf 42} (1979) 873;
  G.~Degrassi, S.~Di Vita, J.~Elias-Mir\'o, J.R.~Espinosa, G.F.~Giudice, G.~Isidori and A.~Strumia,
  JHEP {\bf 1208} (2012) 098
  \arXiv{1205.6497}{hep-ph};
  D.~Buttazzo, G.~Degrassi, P.P.~Giardino, G.F.~Giudice, F.~Sala, A.~Salvio and A.~Strumia,
  JHEP {\bf 1312} (2013) 089
  \arXiv{1307.3536}{hep-ph};
  A.V.~Bednyakov, B.A.~Kniehl, A.F.~Pikelner and O.L.~Veretin,
  Phys.\ Rev.\ Lett.\  {\bf 115} (2015) 20,  201802
  \arXiv{1507.08833}{hep-ph}.

\bibitem{Langer}
J.S.~Langer,
  Annals Phys.\  {\bf 41} (1967) 108
   [Annals Phys.\  {\bf 281} (2000) 941].
  

\bibitem{multi}
  J.R.~Espinosa and T.~Konstandin,
  JCAP {\bf 1901} (2019)  051
  \arXiv{1811.09185}{hep-th}.

\bibitem{CdL}
  S.R.~Coleman and F.~De Luccia,
  Phys.\ Rev.\ D {\bf 21} (1980) 3305.
 
\bibitem{grav}
  J.R.~Espinosa,
  \arXiv{1808.00420}{hep-th}.
   
\bibitem{Russak} 
I.B. Russak, ``Calculus of Variations MA 4311 Lecture Notes'',
Naval Postgraduate School, Monterey, California, USA. 
\url{https://calhoun.nps.edu/handle/10945/39311}.
 
\end{thebibliography}
\end{document}